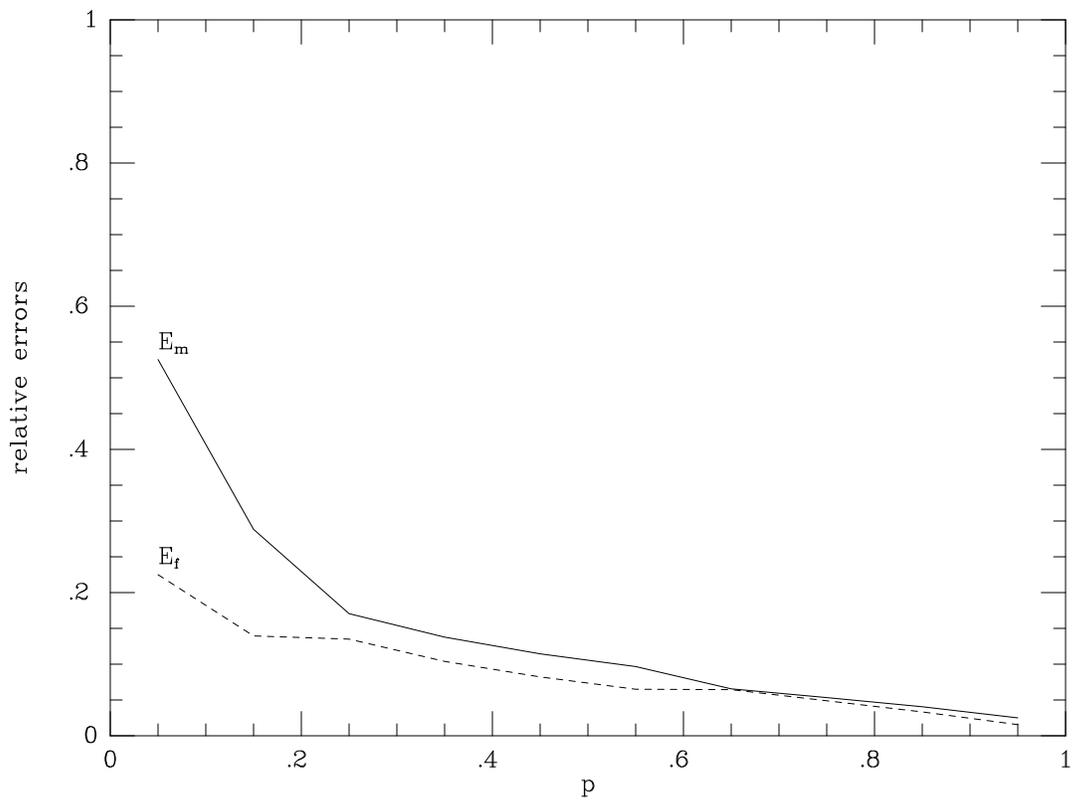

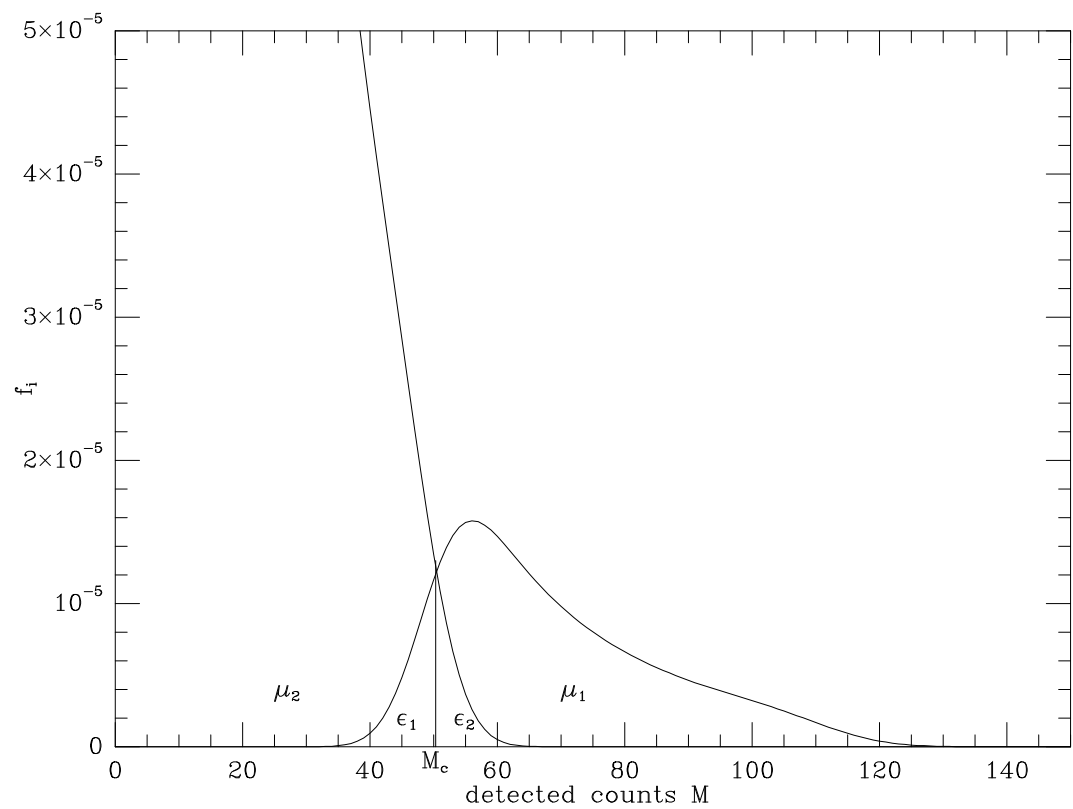

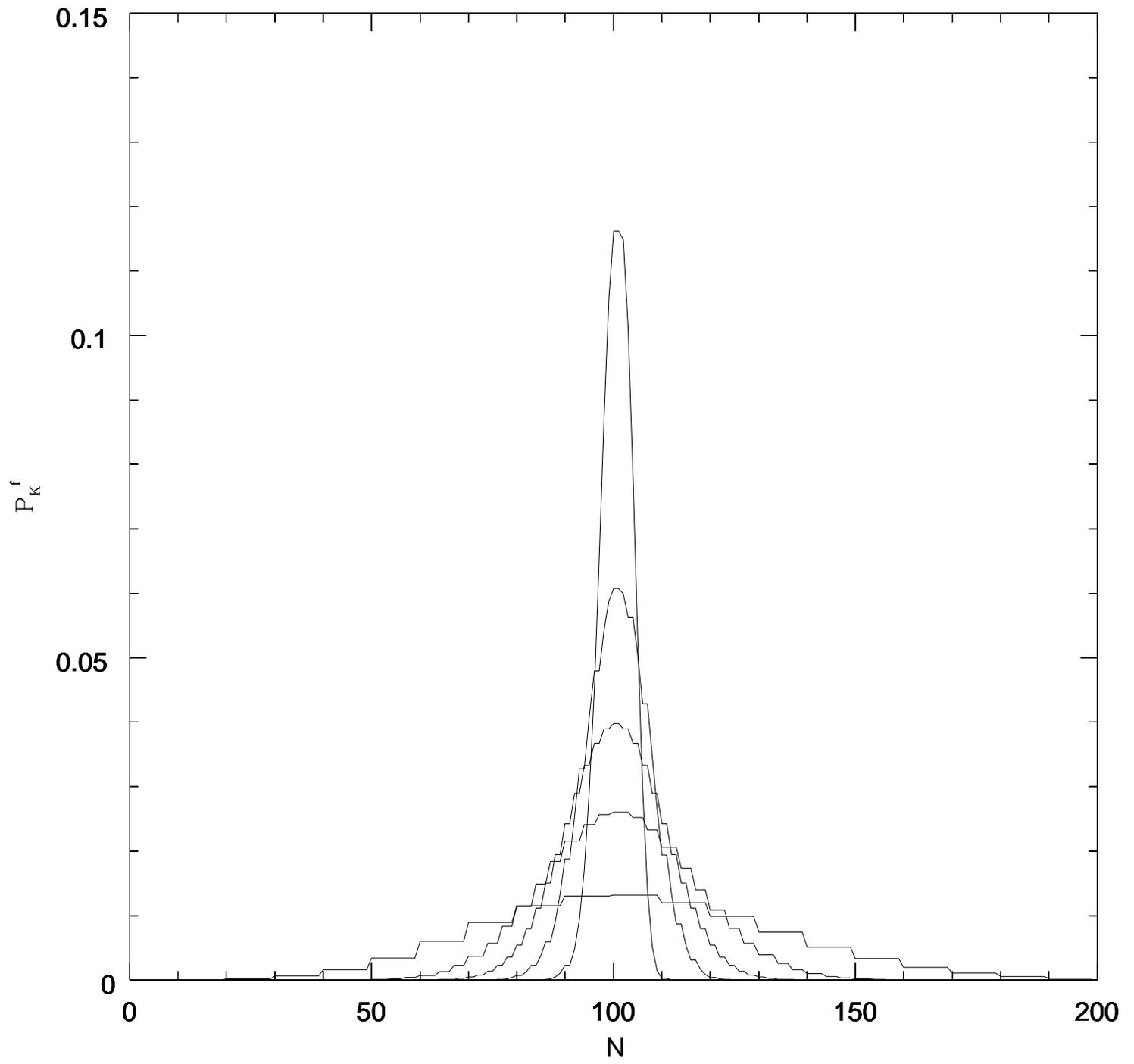

# EFFECTS OF SAMPLING ON MEASURING GALAXY COUNT PROBABILITIES

István Szapudi[1] and Alexander S. Szalay[2,3]

## ABSTRACT

We investigate in detail the effects of sampling on our ability to accurately reconstruct the distribution of galaxies from galaxy surveys. We use a simple probability theory approach, Bayesian classifier theory and Bayesian transition probabilities. We find the best Bayesian estimator for the case of low sampling rates, and show that even in the optimal case certain higher order characteristics of the distribution are irretrievably washed out by sparse sampling: we illustrate this by a simple model for cluster selection. We show that even choosing an optimal threshold, there are nonzero numbers for both misidentified clusters and true clusters missed. The introduction of sampling has an effect on the distribution function that is similar to convolution. Deconvolution is possible and given in the paper, although it might become unstable as sampling rates become low. These findings have important consequences on planning and strategies of future galaxy surveys.

[1] FERMILAB Astrophysics, MS-209, P.O. Box 500 Batavia IL 60510

[2] Dept. of Physics and Astronomy, The Johns Hopkins University, Baltimore, MD 21218

[3] Dept. of Physics, Eötvös University, Puskin U. 5-7. Budapest, Hungary

# 1. INTRODUCTION

Statistical analysis of the Large Scale Structure provides valuable information about both the initial fluctuations in the early Universe, and the subsequent physical processes. The distribution of galaxies can be modeled by a (generalized) Poisson process acting on an underlying continuous random field (Peebles 1980, Szapudi & Szalay 1993a, hereafter SS). The two-point correlation function describes the gross statistical properties of the system. It is analogous to the dispersion for probability distributions. As such, it can only represent the random field well, when it is close to Gaussian, otherwise we need higher order correlation functions for a better description. Recently much effort has been focused on the observational determination of higher order properties of galaxy distributions (eg. Gaztañaga 1992, Meiksin, Szapudi & Szalay 1992, Szapudi, Szalay & Boschán 1992, Bouchet *et al.* . 1993, Szapudi *et al.* 1994, Gaztañaga 1994). As a result, the validity of the hierarchical assumption (Peebles 1980, Balian & Schaeffer 1989) has been established between the higher order correlation functions, implying a strong non-Gaussianity, which is expected from gravitational clustering with Gaussian initial conditions. One of the major obstacles in studying the statistics of this random field is the discrete nature of galaxies: Poisson noise can dilute characteristics that would be immediately visible in a continuum representation. These problems are increasingly prominent when one tries to concentrate on the higher order statistics. Kaiser (1986) proposed that for determining the two point correlation function it is most efficient to use a sparsely sampled redshift survey. For the CfA survey he finds 5% sampling to be optimal for redshifts measured individually, and 10% if the redshifts are obtained with a multiobject-spectrograph. Saunders *et al.* (1991) reconstructed the density field of the local Universe from the QDOT redshift survey which consists of 'one in six' randomly sampled IRAS galaxies. They used generating functions to account for the shot noise from the discreteness of galaxy counts and thereby estimate the moments of the underlying density field. The connection between the continuum and the discrete representations was investigated in a more general setting by SS, and the so called factorial moments emerged as the natural counterpart of the continuum moments. Factorial moments are automatically free from Poisson noise terms, and, as will be shown, they scale with the sampling rate.

In this paper we investigate the effects of different sampling rates in detail, especially for reconstruction of density fields, higher order moments and probability distributions. These findings can be applied when designing galaxy surveys, and for defining cluster finding algorithms. In §2 we present the basic mathematical formalism, §3 expresses the results in terms of Bayesian classifers for cluster finding applications, §4 calculates Bayesian transitional probabilities for density reconstruction, and in §5 we propose some methods for measuring galaxy count probabilities.

## 2. COUNTING EFFICIENCY

Consider $N$ galaxies in a cell, each galaxy detected with a probability $p$. Here $p$ is the product of $\psi$, the selection function for the cell, and the sampling rate. The number of galaxies actually detected, $M$ (we use $M$ for the observed counts throughout) will follow a binomial distribution:

$$P(M \mid N, p) = \binom{N}{M} p^M q^{N-M}, \qquad p + q = 1. \tag{2.1}$$

If the probability of having $N$ galaxies in the cell is $P_N$, the probability of detecting $M$ galaxies will be

$$\bar{P}_M = \sum_N P_N \binom{N}{M} p^M q^{N-M}. \tag{2.2}$$

Throughout this paper we define the binomial coefficents $\binom{N}{M} = 0$ if $M > N$. We can calculate the generating function for the observed distribution:

$$\bar{G}(x) = \sum_{M=0}^{\infty} \bar{P}_M x^M = \sum_{N=0}^{\infty} P_N (q + px)^N = G(q + px) \qquad (2.3)$$

where $G(x)$ is the generating function of the underlying distribution $P_N$. Since the generating function of sampling one of the galaxies is $b(x) = q + px$, the result can be interpreted as the convolution of the two generating functions: $\bar{G}(x) = G \circ b(x)$. This is a special case of a more general result obtained in SS, which states that if the probability of having N clusters in a cell is $P_N$, and each cluster has the probability distribution $c_k$, then the generating function of the counts is the convolution of the corresponding generating functions: $G \circ c(x)$. Note that, if the generating function is a function of $n(x - 1)$, switching to $\bar{P}$ means $n(q + px - 1) = pn(x - 1)$, i.e. the generating function may be obtained just by the usual $n$ to $pn$ substitution, as used by Saunders *et al.* (1991).

## 3. BAYESIAN CLASSIFIERS

The formalism of the previous section can be utilized to describe cluster selection on a sound mathematical basis. In its simplest form clusters are selected via counts exceeding a predetermined threshold. With incomplete sampling, however, we can only set a threshold in the observed counts chosen to minimize the errors, there may be a large scatter in the true counts. In this section we investigate this problem in detail; in particular we propose choosing the threshold optimally to minimize the total probability of missing or misidentifying a cluster: the Bayesian classifier. It is straightforward to modify our method, when the errors must be optimized in a different sense: e.g. we want a clean sample, at the expense of missing a larger fraction of true clusters.

A threshold in the real counts divides the events in two disjoint classes: $N \geq N_c$ (cluster), and $N < N_c$. The probability distribution of the measured counts $\bar{P}_M$ can be uniquely decomposed into two parts: the probability $f_1(M)$, that $M$ is coming from the first class and the probability $f_2(M)$ that it comes from the second class, thus

$$\bar{P}_M = f_1(M) + f_2(M), \qquad (3.1)$$

with an implicit dependence on the preset threshold $N_c$ and the detection probability $p$. The calculation of $f_i(M)$ is straightforward from Eq. (2.1)

$$f_1(M) = \sum_{N \geq N_c} P(M \mid N, p) P_N \qquad (3.2)$$
$$f_2(M) = \sum_{N < N_c} P(M \mid N, p) P_N.$$

If our cluster selection algorithm uses a threshold $M_c$ in the observed counts to decide what a cluster is, it will make two different kinds of error. It can miss a cluster with probability

$$\epsilon_1 = \sum_{M < M_c} f_1(M), \qquad (3.3)$$

or it can detect a false cluster with probability

$$\epsilon_2 = \sum_{M \geq M_c} f_2(M), \qquad (3.4)$$

For completeness we introduce the $\mu_1$ the probability that we detect a cluster correctly, and $\mu_2$ is that we reject one correctly

$$\mu_1 = \sum_{M \geq M_c} f_1(M), \qquad (3.5)$$
$$\mu_2 = \sum_{M < M_c} f_2(M).$$

The meaning of these quantities is clear from Fig.1. We select the threshold $M_c$ to minimize the total error $\epsilon_1 + \epsilon_2$; this is called Bayesian classifier corresponding to $f_1(M_c) = f_2(M_c)$ (Fukanaga 1990; for explanation see Fig. 1. and the next paragraph) We can characterize the relative errors in a sample of clusters selected by this thresholding:

$$E_m = \frac{\epsilon_1}{\epsilon_1 + \mu_1}, \qquad (3.6)$$
$$E_f = \frac{\epsilon_2}{\epsilon_2 + \mu_1},$$

where $E_m$ is the fraction of true clusters missing, end $E_f$ is the fraction of false clusters in the detected sample. The Bayes error used here corresponds to the overall minimalization of the errors. Note, however, that in certain situations a different error optimization may be necessary: one could choose a threshold for instance such as to keep $\epsilon_2$ low to obtain a clean sample at the expense of missing a lot of clusters, i.e. $\epsilon_1$ larger than optimal. It is obvious how to apply our considerations to this case, or cases when there are more than two classes.

To illustrate these ideas we applied the formalism to the selection of the Abell clusters. We used the probability distribution determined from mixed dark matter $N$-body simulations by Klypin *et al.* (1993).

$$h(x) = 0.086 x^{-2.25} \exp(-0.046x),$$
$$P_N \simeq h(N/N_{cl})/(N_{cl}\bar{\xi}), \qquad (3.7)$$

where $N_{cl} = \langle N \rangle \bar{\xi}$, and $\bar{\xi}$ is the average of the correlation function over of the cell fo volume $V$. Szapudi & Szalay (1993b) estimate Abell's sampling rate from a sample of galaxies brighter than $M_v = -18$ at $p = 0.4$, $\langle N \rangle \simeq 0.035$, and $\bar{\xi} \simeq 40$, for a sphere of an Abell radius ($1.5\,\mathrm{h}^{-1}\mathrm{Mpc}$). There is a one-to-one mapping between $N_c$ and corresponding optimal $M_c$ at a given sampling rate. The inverse relation can be applied similarly to find the best corresponding $N_c$ to a certain $M_c$ in the sense of the minimal Bayesian errors. Fig.1. shows the distributions $f_1$ and $f_2$ when $M_c \simeq 50$ (Abell's threshold to select clusters of richness class 1), corresponding to a threshold of $N_c \simeq 100$ in the real counts. This figure illustrates that the condition for minimal errors is $f_1(M_c) = f_2(M_c)$, since if the threshold is moved either left or right from the intersection of the two curves, the errors will increase, similarly to the Maxwell construction.

Fig.2. plots the fractional errors for $N_c \simeq 100$ (chosen such that it would give the appropriate $M_c \simeq 50$ at 40% sampling) as a function of $p$, the sampling rate. As expected, the errors increase with decreasing sampling rate. From Fig.2. we can determine $E_f \simeq 0.1$, $E_m \simeq 0.15$, at $p = 0.4$. This can be compared to the results of Lucey (1983), where he quotes $15 - 25\%$ false detection, and $15 - 30\%$ missed clusters from detailed Monte-Carlo modelling of Abell's selection process. The difference from our results is understandable, since Abell clusters were selected from an angular galaxy distribution, both errors are subject to projection effects, and we only approximated the true probability distribution with one determined from $N$-body simulations, and the sampling rate used is only approximate. The estimated errors are therefore in good agreement with Lucey's findings.

From Fig.2. we can see that the errors start to increase sharply roughly at around $p = 0.3$, this represents a minimum sampling rate necessary to find clusters efficiently, at lower sampling rate the cluster finding algorithm will break down. We found this behaviour in the concrete examples we investigated, and conjecture that this is a quite generic behaviour, i.e. there exist a sampling under which the errors increase steeply. The exact location of the threshold, however depends on the parameters in a quite complicated way, and the minimal sampling rate is not uniquely defined: it has to be determined in every situation individually keeping in mind the objectives of the survey.

## 4. BAYESIAN TRANSITION PROBABILITES, DENSITY RECONSTRUCTION

Often we need to extract statistical information from a sample beyond simply knowing whether the counts are above or below a threshold. When measuring the count $M$ the question naturally arises as to how good a constraint this measurement imposes on the underlying distribution. Using Bayes' theorem we calculate the inverse probability $P(K|M,p)$, the probability distribution of the real counts, $K$, given the measured $M$ as

$$P(K|M,p) = \frac{P(M|K,p)P_K}{\bar{P}_M} = \frac{\binom{K}{M}q^K P_N}{\sum_{S=M}^{\infty} \binom{M}{S}q^S P_S}. \qquad (4.1)$$

Previously we calculated $P(M|N,p)$, the probability that our measurement will yield $M$ if the real count is $N$. Now an interesting picture arises: there may be $N$ galaxies in reality, we measure $M$ and *infer* $K$ from this information. We can determine the transition probability between the real counts and our (best possible) inferences:

$$P(K|N) = \sum_M P(K|M)P(M|N) = P_K \sum_M \frac{\binom{N}{M}\binom{N}{K}p^M q^{N+K-M}}{\sum_S P_S \binom{S}{M}q^S}. \qquad (4.2)$$

In the case of full sampling this function is simply a discrete Dirac (or a Kronecker) delta. As we lower the sampling rate, the Dirac delta broadens into a function that represents the uncertainty we have when reconstructing the real distribution from the measured counts, and for a given sampling rate it also depends on the underlying $P_N$.

With this transition probability we can easily calculate using Eq. (4.2) the inferred probability distribution $\hat{P}_K$

$$\hat{P}_K = \sum_N P(K|N)P_N = P_K. \qquad (4.3)$$

This means, that although our inference might be smeared out, it is unbiased in the sense of the whole distribution (i.e. we can still accurately reconstruct the probability distribution as a whole). This is a non-trivial equation, because the kernel is not a delta function.

Let us consider the naive estimate, $K = M/p$, where $M$ is the measured value at sampling $p$, and $K$ is the more usual inference arising simply by divison with the sampling. This corresponds to a $P(K|M,p) = \delta_{M,Kp}$, or uniform priors in Bayesian language. The inferred distribution $\hat{P}_K^f$ is

$$\hat{P}_K^f = \sum_N \binom{N}{Kp} p^{Kp} q^{N-Kp} P_N. \tag{4.4}$$

This clearly cannot reconstruct the distribution under low sampling rates. Even if the distribution is a sharp delta, $P_N = \delta_{NN_0}$, our inference will be smeared out as $\hat{P}_K^f = \binom{N_0}{Kp} p^{Kp} q^{N_0-Kp}$. This is demonstrated on Fig. 3. which shows how the response of this approach to a sharp delta broadens with decreasing sampling rate. Since any distribution can be decomposed as the sum of delta functions, this argument shows, that the naive estimate cannot reconstruct a distribution from infinite measurements.

## 5. DETERMINING THE TRUE $P_N$ DISTRIBUTION

Section §3 illustrated in conjunction with cluster selection, that the information irretrievably lost through incomplete sampling cannot be recovered even with Bayesian approach. The previous sections showed, however, that Bayesian methods reconstruct the density in a statistical sense (we illustrated this with counts in cells), while the simple approach of dividing with the sampling does not. For these theoretical calculations we assumed that the galaxy count probabilities are a priori known from measurements or theory. Without the aim of completeness this section contains some suggestions and ideas about how such measurements can be performed using one-point informations only. This puts the the previous considerations in a more practical framework. Detailed evaluation and comparison of these models together with other higher order methods to be used for density reconstruction is left for subsequent research.

For the case of a catalog with uniform sampling rate, like galaxies on a radial shell, we can explicitly express the inversion. We calculated $\bar{G}(x) = G \circ b(x)$ before. If we define $b^{-1}(x)$ such that $b \circ b^{-1} = \text{id}$, or explicitly $b^{-1}(x) = x/p - q/p$, we can invert the formula for the generating functions, and thus calculate $P_N$ from the measurement of $\bar{P}_M$ as

$$P_N = \sum_{M=N}^{\infty} \binom{M}{N} p^{-M} (-q)^{M-N} \bar{P}_M. \tag{5.1}$$

This equation is always properly normalized, as can be seen from taking the generating function at $x = 1$. The result is a correct mathematical inversion, so as long as the underlying probability distribution is positive definite, the reconstruction from a perfect measurement will yield positive definite results. If the measurement is finite, but a good approximation, the result will still be a good approximation of the original distribution and as such positive. In a practical situation, if negative values are encountered, they simply mean the lack of adequate information to reconstruct the the probability distribution. The application of this formula (and any similar inversion method for that matter) is limited by the fact that it is an alternating sum, which is subject to instabilities at low sampling rates, especially in the tail of the distribution, where a small numbers are determined by subtracting large numbers from each other. The probability of $N > N_0$, (probability of overdensity) can be inverted similarly. If $M$ galaxies are measured, the result can be obtained from the sum of Eq. (2.1) for all $N \geq N_0$

$$P(N \geq N_0 | M) = \frac{\sum_{N=N_0}^{\infty} \binom{N}{M} q^N P_N}{\sum_{S=M}^{\infty} \binom{M}{S} q^S P_S}. \tag{5.2}$$

Unfortunately, in real (astrophysical) situations the selection function might change from place to place, it might be different for each bin in which we measure the counts. In that case the former inversion breaks down because in principle there are many distributions that could produce the same data. One possibility is to take a maximum likelihood approach: what distribution is most likely to produce the measurements? If we determine this distribution with enough measurements, it will be a good approximation to the real underlying distribution. Let us index our bins with $i$ from 1 to $C$ and assume that at bin $i$ where the probability is $p_i$, we measure $M$. According to the previous consideration we can determine $\bar{P}_i(M_i)$ which will depend on all the $P_N, N \geq M$ and the selection function at that bin. Since our measurements are independent we can construct the probability $\bar{P}(M_1, \ldots M_C)$ which will be the product of the probabilities for each bin:

$$\bar{P}(M_1, \ldots M_C) = \prod_{i=1}^{C} \sum_{N_i=M_i}^{\infty} P_{N_i} \binom{N_i}{M_i} p_i^{M_i} q_i^{N_i - M_i} \tag{5.3}$$

We can maximize this expression as a function of $P_N$ to get the best estimator for the underlying distribution. When all the selection functions are constant the maximization can be done analytically, and it gives back the usual estimator for $P_N$: the number of bins with $N$ galaxies divided by the sum of all bins. The general case can be handled numerically.

Another possibility is to characterize a distribution by its moments, and use the scaling properties of the moments for inversion. One might naively expect that the moments scale with the sampling rate $N^k \simeq p^k$. A closer look at the problem reveals that it is not the case. Let us introduce the notation $(N)_k = N(N-1)\ldots(N-k+1)$, and define the factorial moments as

$$\langle (N)_k \rangle = \sum_{N \geq 0} P_N (N)_k \tag{5.4}$$

The defined factorial moments scale with $p$ because

$$\langle (M)_k \rangle = \frac{d^k}{dx^k}|_{x=1} \bar{G}(x) = p^k \frac{d^k}{dx^k}|_{x=1} G(x) = p^k \langle (N)_k \rangle \tag{5.5}$$

This could be anticipated because (according to SS) the factorial moments (and not the regular moments) of the discrete distribution are equal to the continuum moments of the underlying field, and the continuum moments clearly scale with the sampling. Note that the factorial moments are equivalent to the conventional method of subtracting shot noise contribution, however, the results are simpler and scale with the sampling rate. For the sake of completeness we cite the formula from SS for the connection of the factorial moments with the ordinary moments:

$$\langle N^m \rangle = \sum_{k=0}^{m} S(m,k) F_k, \tag{5.6}$$

where $S(m,k)$ are the Stirling numbers of the second kind.

These results define an estimator $F_k$ for the true factorial moments of the distribution, provided that the selection probability $p_i$ is known for each cell:

$$F_k = \frac{1}{C} \sum_i \frac{(N_i)_k}{p_i^k}, \tag{5.7}$$

where $N_i$ is the count in cell $i$. This is an unbiased consistent estimator for the factorial moments. However, the errors of the estimates for the selection probability can introduce further errors and instabilities, especially at low sampling rates. Once we know the factorial moments at sufficient accuracy, it is a simple matter to find the probability distribution using that the generating function of the probability distribution $G(x)$ can be written in terms if the exponential generating function of the factorial moments $F(x) = \sum_k F_k \frac{x^k}{k!}$ as

$$G(x) = F(x-1) \tag{5.8}$$

(see SS for details). The inversion calculated this way will be by definition unbiased in the sense of factorial moments. The application of the expansion of this equation is limited by the same restrictions as mentioned after Eq. (5.1).

Another possibility would be to artificially 'dilute' the sample by randomly selecting a fraction of galaxies in the more sampled areas, such that the sampling will be constant in the resulting survey. By doing this several times, it can be ensured each galaxy will be in one of the resulting catalogs at least once statistically. After calculating the average probability distribution over this ensemble of catalogs Eq. (5.2) is dierctly applicable.

## 6. DISCUSSION

We investigated the sampling effects on the reconstruction of galaxy counts in cells. One of our examples addresses the problem of cluster selection using a simple model: we wish use a galaxy catalog which samples the the galaxies at a constant (possibly low) rate to select locations of the fully sampled catalog exceeding a preset threshold $N_c$. We find that there is an optimal threshold in the undersampled catalog ( $M_c \simeq N_c p$, where $p$ is the sampling rate, *approximately* for high sampling rates, but not exactly) that minimizes the total errors. This minimal error corresponds to the so called Bayesian classifier and constitutes one of the main results of the paper. If the desired theshold $N_c$ is fixed, the only possiblity to lower these errors is to increase the sampling rate. This shows, that, as expected, information is irretreivably lost about the actual density map through incomplete sampling. The formalism for calculating the Bayesian classifer, which minimizes the total error, can also be used when different error optimization is needed, and can be generalized even for multiple classes of objects, e.g. clusters of different richness classes. As an example, we applied these considaration to the selection of Abell clusters, obtaining results for the fractions of missing true clusters, and misidentified clusters in agreement with previous findings.

We studied the reconstruction of the distribution of the count probabilities as well. We found that under ideal conditions of large number of (independent) meaurements and fixed sampling rate, the Bayesian inversion provides an unbiased estimator of the count probabilities in the sense of the moments of the distribution, while the naive approach (with uniform priors) smears out the distribution substantially under the same conditions. If the sampling rate is low: it cannot even recover a sharp (delta function) distribution from infinite measurements. The case of variable sampling is more complicated, and requires an even more sophisticated approach; we proposed several of these. The distribution, according to the previous findings, can only be recovered in the statistical sense, since part of the information is lost through incomplete sampling. Finally, note, that in this paper we did not deal with the 'cosmic' errors for the reconstruction of the probability distribution arising from the fact that a real galaxy catalog contains only a finite portion of the Universe. This is a related, but nevertheless different problem, for which a solution will be presented in Szapudi & Colombi (1995).

The authors would like to thank the referee, E. Gaztañaga for suggested improvements. This research has been supported by an NSF/Hungary exchange grant, by a grant from the US-Hungary Joint Fund, by NSF grant AST 90-20380 and by the Seaver Foundation. IS was supported by DOE and by NASA through grant NAG-5-2788 at Fermilab.

# REFERENCES

Balian, R. & Schaeffer, R. 1989, A&A, 220, 1

Bardeen, J.R., Bond, J.R., Kaiser, N. & Szalay, A.S. 1986, ApJ, 304, 15

Bouchet, F. *et al.* 1993, ApJ, 400, 25

Fukanaga, S. 1994, Introduction to Statistical Pattern Recognition (San Diego: Academic Press)

Gaztañaga, E. 1992, ApJ, 398, L17

Gaztañaga, E. 1994, MNRAS, 268, 913

Kaiser, N. 1986, MNRAS, 219, 785

Klypin. A.A., Holtzman, J., Primack, J., & Regös, E. 1993, ApJ, 416, 1

Lucey, J. 1983, MNRAS, 204, 33

Meiksin, A., Szapudi, I. & Szalay, A.S. 1992, ApJ, 394, 87

Peebles, P.J.E. 1980, The Large Scale Structure of the Universe (Princeton: Princeton Univ. Press)

Saunders, W. *et al.* 1991, Nature, 349, 32

Szapudi, I., Szalay, A.S., & Boschán P. 1992, ApJ, 390, 350

Szapudi,I. & Szalay, A.S. 1993a, ApJ, 408, 43 (SS)

Szapudi,I. & Szalay, A.S. 1993b, ApJ, 414, 493

Szapudi,I. & Colombi, S. 1995, ApJ, in preparation,

Szapudi,I., Dalton G., Szalay, A.S. and Efstathiou, G.P. 1995, ApJ, 444, 520

FIGURE CAPTIONS

FIG.1.— The distributions $f_1$ and $f_2$ are shown. $M_c \simeq 50$ (Abell's threshold to select clusters of richness class 1), corresponding to a threshold of $N_c \simeq 100$ in the real counts, is at the intersection of the two curves. The different contributions to the error and detection probabilities are the areas under the curves, denoted by $\epsilon_1$, $\epsilon_2$, $\mu_1$, $\mu_2$ (see text for details).

FIG.2.— The fractional errors for false detection $E_f$, and for missed clusters $E_m$ are shown for $N_c \simeq 100$ as a function of $p$, the sampling probability. As expected, the errors increase with decreasing sampling rate. For Abell's estimated 40% sampling rate we can determine $E_f \simeq 0.1$, $E_m \simeq 0.15$.

FIG.3.— The response of the 'naive' approach to the a sharp (Kronecker-delta) distribution is displayed with sampling rates $p = 0.1, 0.3, 0.5, 0.7, 0.9$. The resulting distributions broaden with decreasing sampling rate; only full sampling would recover the delta function. Theoretically the Bayesian approach would recover the delta function at any sampling rate.